\documentclass[fleqn,usenatbib]{mnras}
\usepackage{amsmath}
\usepackage{siunitx}
\usepackage{multirow}
\usepackage{graphicx}
\usepackage{pifont}

\newcommand{\xmark}{\ding{55}}
\newcommand{\cmark}{\ding{51}}
\DeclareSIUnit\parsec{pc}
\DeclareSIUnit\year{yr}
\DeclareSIUnit\erg{erg}
\DeclareSIUnit\photon{photon}
\DeclareSIUnit\Msun{M_{\odot}}
\title[Test of SGRB jet \& delay models]{A global test of jet structure and delay time distribution of short-duration gamma-ray bursts}
\author[J-W Luo et al.]{Jia-Wei Luo$^{1,2}$\thanks{luoj7@unlv.nevada.edu},
	 Ye Li$^{3}$\thanks{yeli@pmo.ac.cn},
	 Shunke Ai$^{1,2}$,
	 He Gao$^{4}$,
	 and Bing Zhang$^{1,2}$\thanks{bing.zhang@unlv.edu}
\\
$^{1}$Nevada Center for Astrophysics, University of Nevada, Las Vegas, NV 89154, USA\\
$^{2}$Department of Physics and Astronomy, University of Nevada, Las Vegas, NV 89154, USA\\
$^{3}$Purple Mountain Observatory, Chinese Academy of Sciences, Nanjing 100012, China\\
$^{4}$Department of Astronomy, Beijing Normal University, Beijing 100875, China}

\date{Accepted XXX. Received YYY; in original form ZZZ}
\pubyear{2022}
\begin{document}
\label{firstpage}
\pagerange{\pageref{firstpage}--\pageref{lastpage}}
\maketitle

\begin{abstract}
	The multi-messenger joint observations of GW170817 and GRB170817A shed new light on the study of short-duration gamma-ray bursts (SGRBs). Not only did it substantiate the assumption that SGRBs originate from binary neutron star (BNS) mergers, but it also confirms that the jet generated by this type of merger must be structured, hence the observed energy of an SGRB depends on the viewing angle from the observer. However, the precise structure of the jet is still subject to debate. Moreover, whether a single unified jet model can be applied to all SGRBs is not known. Another uncertainty is the delay timescale of BNS mergers with respect to star formation history of the Universe. In this paper, we conduct a global test of both delay and jet models of BNS mergers across a wide parameter space with simulated SGRBs. We compare the simulated peak flux, redshift and luminosity distributions with the observed ones and test the goodness-of-fit for a set of models and parameter combinations. Our simulations suggest that GW170817/GRB 170817A and all SGRBs can be understood within the framework of a universal structured jet viewed at different viewing angles. Furthermore, models invoking a jet plus cocoon structure with a lognormal delay timescale is most favored. Some other combinations (e.g. a Gaussian delay with a power-law jet model) are also acceptable. However, the Gaussian delay with Gaussian jet model and the entire set of power-law delay models are disfavored.
\end{abstract}


\maketitle

\section{Introduction}
\label{sec:introduction}

Gamma-ray bursts (GRBs), as the most energetic transient events in the Universe, are generally categorized into two types based on their durations \citep{kouveliotou1993IdentificationTwoClasses} and multi-wavelength observational criteria \citep{zhang2009DISCERNINGPHYSICALORIGINS,li2016ComparativeStudyLong}. Long GRBs (LGRBs) are believed to be originated from core-collapse of massive stars \citep{woosley1993GammarayBurstsStellar}, while short GRBs (SGRBs) are deemed to be the result of compact star mergers \citep{eichler1989NucleosynthesisNeutrinoBursts}. The core-collapse model for LGRBs is supported by direct observational evidence of the association of some LGRBs with Type Ic supernovae \citep{galama1998UnusualSupernovaError, woosley2006SupernovaGammarayBurst}. The compact star merger model for SGRBs, on the other hand, has only been supported through indirect evidence such as host galaxy type or position of GRBs within the host galaxies \citep{gehrels2005ShortGrayBurst,nakar2007ShorthardGammarayBursts,berger2014ShortdurationGammarayBursts}.

In August 2017, the multi-messenger electromagnetic and gravitational wave observations of GW170817/GRB 170817A originated from a binary neutron star merger event \citep{abbott2017GW170817ObservationGravitational,abbott2017MultimessengerObservationsBinary,abbott2017GravitationalWavesGammaRays} provided clear evidence for the compact star merger origin of SGRBs \citep{abbott2017GW170817ObservationGravitational,goldstein2017OrdinaryShortGammaRay,zhang2018PeculiarLowluminosityShort}. This SGRB is special in its relatively weak prompt emission and its peculiar afterglow lightcurve, which suggests that the jet must possess some type of structure and the GRB is viewed off-axis \citep{troja2017XrayCounterpartGravitationalwave,xiao2017AfterglowsKilonovaeAssociated,zhang2018PeculiarLowluminosityShort,lazzati2018LateTimeAfterglow,lyman2018OpticalAfterglowShort,troja2019YearLifeGW170817,beniamini2019LessonGW170817Most, troja2020ThousandDaysMerger,cheng2021AfterglowEmissionStratified}.

Some structured jet models have been proposed, including the power-law and Gaussian jet models \citep{zhang2002GammaRayBurst,rossi2002AfterglowLightCurves} and two-component models invoking a central jet and a surrounding cocoon \citep{zhang2004PropagationEruptionRelativistic}. These models were originally discussed within the context of LGRBs, but were also tested against SGRBs after the discovery of GW170817/GRB 170817A.  The current observational data from GW170817 alone is not sufficient to directly differentiate among different models \citep{nakar2018ImplicationsRadioXray,troja2020ThousandDaysMerger,oganesyan2020StructuredJetsXRay,takahashi2021DiverseJetStructures}. Some studies \citep{beniamini2019ObservationalConstraintsStructure,salafia2020GammarayBurstJet,hayes2020ComparingShortGammaRay,guo2020LuminosityDistributionShort,takahashi2020InverseReconstructionJet,beniamini2020AfterglowLightcurvesMisaligned,tan2020JetStructureIntrinsic,lloyd-ronning2020ConsequencesGammarayBurst,dado2022CriticalTestsLeading,urrutia2021WhatDeterminesStructure,preau2021NeutronConversionDiffusion,tandon2021UnderstandingGammarayBurst,beniamini2022RobustFeaturesOffAxis,rhodes2022JetCocoonGeometryOptically} have confronted various structured jet models against the data and constrained model parameters. However, it is not known whether the structured jet model constrained from the GW170817/GRB 170817A data can be applied to the entire SGRB population. This is an important question, which is related to whether SGRBs have a {\em quasi-universal} structured jet, with  SGRBs with different properties being largely due to different viewing angles with respect to the jet axis. Such an elegant picture has been widely tested against the data of active galactic nuclei \citep{urry1995UNIFIEDSCHEMESRADIOLOUD} and LGRBs \citep{zhang2002GammaRayBurst,rossi2002AfterglowLightCurves,perna2003JetsGammaRay,dai2005GlobalTestQuasi}.

One intriguing feature of GRB 170817A/GW170817 that might be related to its jet structure is the observed $\sim\SI{1.7}{\second}$ delay time between the binary merger and the gamma-ray burst \citep{abbott2017MultimessengerObservationsBinary,goldstein2017OrdinaryShortGammaRay,zhang2018PeculiarLowluminosityShort}. \citet{zhang2019DelayTimeGravitational} suggested that this delay time $\Delta t$ consists of three components: jet launching waiting time $\Delta t_{\rm jet}$, jet break-out time $\Delta t_{\rm bo}$ and the time for the jet to reach gamma-ray emission radius $\Delta t_{\rm GRB}$. Some numerical simulations of jet propagation typically assumed $\Delta t_{\rm jet}$ of the order of $\sim 1$s \citep[e.g.][]{gottlieb2018CocoonEmissionElectromagnetic}. The launched jet therefore undergoes significant interactions with the ejecta and producing a prominent cocoon surrounding the jet.  \cite{geng2019PropagationShortGRB} performed a series of numerical simulations of jet propagation into the surrounding the ejecta, paying special attention to the effect of $\Delta t_{\rm jet}$. They found that the jet structure indeed depends on this parameter. The prominent cocoon structure is only apparent when $\Delta t_{\rm jet}$ is longer enough (e.g. $>0.5 s$). For a shorter $\Delta t_{\rm jet}$, the jet quickly penetrates through the surrounding ejecta and develops its own structure because of the loss of confinement. The jet structure is close to a Gaussian shape without significant cocoon component surrounding the jet. Thus identifying jet structure observationally may help one to diagnose $\Delta t_{\rm jet}$ and the jet launching physics currently poorly understood. 

Another important question for binary neutron star (BNS) mergers is regarding their delay with respect to star formation. It is known that BNS mergers must be delayed, but the distribution of the delay timescale $\tau$ is not well constrained. Earlier models suggested that the distribution should follow a power-law \citep[e.g.][]{nakar2006LocalRateProgenitor}. However, a later systematic study of a SGRB sample suggested that this simple model cannot reproduce the data. Rather, one requires a characteristic delay timescale, with the distribution either in the Gaussian or lognormal forms \citep{virgili2011AREALLSHORTHARD,wanderman2015RateLuminosityFunction}. It is unclear how the current, much larger SGRB sample confronts with these delay models. 

The observed SGRB population is a convolution of their luminosity function, which depends on the jet structure within the quasi-universal jet hypothesis, and their redshift distribution, which depends on the delay timescale distribution models. It is desirable to perform a global test across different models and parameter spaces against the SGRB and GW170817/GRB 170817A data. This is the purpose of this paper. In Section \ref{sec:method}, we present our Monte Carlo simulations of SGRB populations by considering three different delay time distribution models and three structured jet models. In Section \ref{sec:test_models_with_data} these nine models are confronted by two SGRB samples: one $z$-known sample (mostly dominated by the Swift population) and another $z$-unknown sample (mostly dominated by the Fermi GBM population). The results are reported in Section \ref{sec:results} and the conclusions are drawn in Section \ref{sec:conclusions}.

\section{Models for SGRB simulation}
\label{sec:method}

\subsection{Delay time distribution models}

SGRBs are believed to be originated from neutron star mergers, either NS-NS or NS-BH mergers. These mergers take an extra time with respect to star formation to allow the two compact objects to form and more importantly, to allow the binary to lose orbital energy and angular momentum via gravitational wave radiation before merging \citep{faber2012BinaryNeutronStar,burns2020NeutronStarMergers}. 

If we assume that the fraction of mass to form compact binary systems over all the mass to form new-born stars remains constant in cosmological time, then the SGRB rate density $R_{SGRB}$ (in units of \si{\per\mega\parsec\cubed\per\giga\year}) can be estimated as a convolution of star formation rate density (SFRD) $\phi$ (in units of \si{\Msun\per\mega\parsec\cubed\per\giga\year}) and the probability density function of delay time distribution $f(\tau)$ \citep{sun2015EXTRAGALACTICHIGHENERGYTRANSIENTS}:

\begin{equation}
R_{\rm SGRB}(z) = \int_{\tau_{min}}^{\tau_{max}} \phi[z'(\tau)]f(\tau) \,\mathrm{d}\tau,
\end{equation}
where $z'(\tau)$ is the redshift at the formation of the binary star system while $z$ is the redshift of the SGRB event. The delay time between star formation and merger is $\tau = t(z)-t(z')$, where $t(z)$ and $t(z')$ are the age of universe at redshifts $z$ and $z'$, respectively. Here $\phi[z'(\tau)]$ is star formation rate density (SFRD) at the formation of the binary star system. 

Since the progenitors of neutron stars usually have $\SIrange{10}{29}{\Msun}$ initial mass, and the lifetimes of them are $\SIrange{10}{30}{\mega\year}$. A minimum delay time $\tau_{min} = \SI{10}{\mega\year}$ is required. The age of universe at redshift $z$, $t(z)$ is used as $\tau_{max}$. We use Equation 5 of \citet{yuksel2008RevealingHighredshiftStar} as SFRD $\phi(z)$:
\begin{equation}
\phi(z) = \rho_0\left[(1+z)^{a\eta}+\left(\frac{1+z}{B}\right)^{b\eta}+\left(\frac{1+z}{C}\right)^{c\eta}\right]^{1/\eta}
\end{equation}
Where $\rho_0=0.02$, $a=3.4$, $b=-0.3$, $c=-3.5$, $\eta=-10$, $B=5000$, $C=9$.

\begin{figure}
	\includegraphics[width=0.49\textwidth]{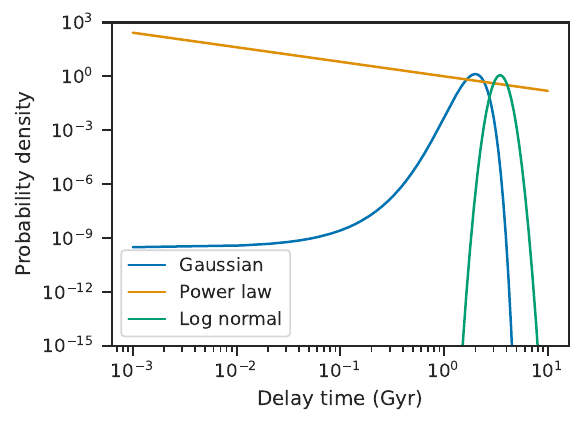}
	\caption{Probability density distributions of different delay functions assuming normalization to unity. Gaussian delay: $t_G=2.0, \sigma_G=0.3$. Power-law delay: $\alpha=0.81$. Lognormal delay: $t_{LN}=3.5,\sigma_{LN}=0.1$.}
	\label{fig:delaymodel}
\end{figure}

As for the delay time distribution, we consider the following three function forms that have been discussed in the literature, with examples shown in Fig. \ref{fig:delaymodel}:

\begin{enumerate}
	\item Gaussian:
	\begin{equation}
	f(\tau) \propto \exp\left(-\frac{(\tau-t_G)^2}{2\sigma_{\rm G}^2}\right)/\left(\sqrt{2\upi}\sigma_{\rm G}\right).
	\end{equation}
	A previous analysis suggested $t_{\rm G}=\SI{2}{\giga\year}$ and $\sigma_{\rm G}=\SI{0.3}{\giga\year}$ \citep{virgili2011AREALLSHORTHARD}.
	
	\item Lognormal:
	\begin{equation}
	f(\tau) \propto \exp\left(-\frac{(\ln\tau-\ln t_{LN})^2}{2\sigma_{LN}^2}\right)/\left(\sqrt{2\upi}\tau\sigma_{LN}\right).
	\end{equation}
	A previous analysis suggested $t_{LN}=\SI{2.9}{\giga\year}$ and $\sigma_{LN}=0.2~{\rm ln(Gyr)}$ \citep{wanderman2015RateLuminosityFunction}.
	
	\item Power-law
	\begin{equation}
	f(\tau) \propto \tau^{-\alpha}.
	\end{equation}
    Such power-law model is supported by the study of host galaxy mass distribution and stellar age distributions \citep{berger2014ShortdurationGammarayBursts}. It is also consistent with the merger time distribution of Galactic NS-NS binaries \citep{piran1992ImplicationsComptonGRO}.	By comparing the ratio between early-type host galaxies and late-type host galaxies, \cite{zheng2007DeducingLifetimeShort} suggested that $\alpha=1$.  On the other hand, \cite{wanderman2015RateLuminosityFunction} suggested $\alpha=0.81$, even though this model is not as good as the lognormal model to interpret the SGRB data.
\end{enumerate}

With the redshift-dependent SGRB rate density specified, the number of SGRBs observed during an observation time duration $T$ within redshift range $z\text{--}\mathrm{d}z$ can be derived as
\begin{equation}
\mathrm{d}N_{\rm SGRB}(z)=R_{\rm SGRB}(z)\frac{T}{1+z}\frac{\mathrm{d}V}{\mathrm{d}z}{\mathrm{d}z},
\end{equation}
where
\begin{equation}
\frac{\mathrm{d}V(z)}{\mathrm{d}z}=\frac{4\upi D_L^2 c}{(1+z)^2H_0}\left[\Omega_m(1+z)^3+\Omega_k(1+z)^2+\Omega_{\Lambda}\right]^{-1/2}
\end{equation}
is the comoving volume of the universe at redshift $z$, $D_L$ is the luminosity distance, $c$ is speed of light. We take the following Planck cosmological parameters in our calculations:  $H_0=\SI{67.66}{\kilo\meter\per\second\per\mega\parsec}$, $\Omega_m=0.3111$, $\Omega_k=0$ and $\Omega_{\Lambda}=0.6889$ \citep{aghanim2020Planck2018Results}. In our simulations, we do not need to consider the normalization of the delay time distribution function but vary the number of SGRBs we simulate to ensure the simulated samples are comparable to the observed samples.

\subsection{Structured jet models}

After simulating the distances of SGRB sources, we then simulate the luminosity distribution of the SGRBs within the framework of a quasi-universal structured jet model. For each structured jet model, we assume that all SGRBs have the same jet structure, so that the angular peak luminosity per solid angle, which is the proxy of the isotropic equivalent peak luminosity, depends on the viewing angle $\theta$. 

Assuming that the jet structure has a rotational symmetry around the jet axis, the angular peak luminosity per solid angle as a function of viewing angle $\theta$ is expressed as $l(\theta)$. For SGRB jets with random orientations, the viewing angle should be isotropically distributed. So in our simulations, we can draw $\sin\theta$ randomly from a uniform distribution and deduce $\theta$ from it. Then for a viewing angle $\theta$, the observed isotropic peak luminosity is 
\begin{equation}
L_{p,iso} = 4\upi l(\theta).
\end{equation}
For convenience, hereafter we will use $L$ to denote the isotropic peak luminosity $L_\mathrm{p,iso}$.

\begin{figure}
	\includegraphics[width=0.49\textwidth]{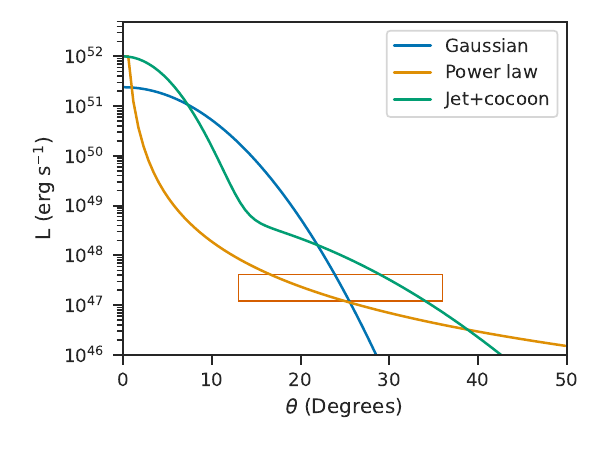}
	\caption{Angular luminosity distributions of different jet models. Gaussian jet: $L_0=2.4\times10^{51} \ {\rm erg \ s^{-1}}, \theta_0=0.1$. Power-law jet: $L_0=1.0\times10^{52} \ {\rm erg \ s^{-1}}, \theta_0=0.01, k=3.0$. Jet+cocoon jet: $L_0=1\times10^{52} \ {\rm erg \ s^{-1}}, \theta_0=0.06, L_1=1\times10^{49} \ {\rm erg \ s^{-1}}, \theta_1=0.2$. The constraint on jet parameters from GW170817 are shown as a orange rectangle.}
	\label{fig:jetmodel}
\end{figure}

There are three main structured jet models discussed in the literature, with examples shown in Fig. \ref{fig:jetmodel}:

\begin{enumerate}
	\item Gaussian jet model \citep{zhang2002GammaRayBurst}
	\begin{equation}
	l(\theta) = l_0 e^{\frac{-\theta^2}{2\theta^2_0}},
	\end{equation}
	where $l_0$ is the maximum luminosity density when the jet is pointed directly at the observer. When the viewing angle $\theta$ is larger than $\theta_0$, the isotropic peak luminosity $L$ falls rapidly. 
	
	\item Power-law jet model \citep{meszaros1998ViewingAngleEnvironment,rossi2002AfterglowLightCurves,zhang2002GammaRayBurst}
	\begin{equation}
	l(\theta)=
	\begin{cases}
		l_0 & \theta<\theta_0, \\
		l_0(\theta/\theta_0)^{-k} & \theta\geq\theta_0, \\
	\end{cases}
	\end{equation}
	where $l_0$ is the energy density at or smaller than a small characteristic angle $\theta_0$. When $\theta > \theta_0$, the angular peak luminosity decreases as a power-law. Note that $l_0$ is needed to avoid divergence to infinity at small angles.
	
	\item Two-component Gaussian jet+cocoon model \citep{bromberg2011PROPAGATIONRELATIVISTICJETS}
	\begin{equation}
	l(\theta) = l_0 e^{\frac{-\theta^2}{2\theta^2_0}} + l_1 e^{\frac{-\theta^2}{2\theta^2_1}}.
	\end{equation}
	This model consists of two Gaussian components: a narrower, brighter jet defined by $l_0$ and $\theta_0$, and a wider, fainter cocoon defined by $l_1$ and $\theta_1$. 
\end{enumerate}

Since we are testing a quasi-universal model for all SGRBs including GRB 170817A, all our parameters are required to reproduce the observed luminosity and viewing angle of GRB 170817A. The constraints from GRB 170817A are shown as an orange rectangle in Fig. \ref{fig:jetmodel}, which represents the uncertainty ranges of $\SI{1.2e47}{\erg\per\s}<L<\SI{4.1e47}{\erg\per\s}$ and $\SI{13.18}{\degree}<\theta<\SI{36.10}{\degree}$ \citep{zhang2018PeculiarLowluminosityShort,troja2020ThousandDaysMerger}, we require the luminosity distributions calculated from the parameter combinations intercept with this rectangle.

\section{Test models against data}
\label{sec:test_models_with_data}

For an easy comparison with the observed values, we will report the jet central luminosity parameter with isotropic peak luminosity $L$ instead of luminosity per solid angle $l$ in the rest of the paper (e.g. $L_0 = 4\upi l_0$ and $L_1 = 4\upi l_1$ in Tables \ref{tab:parameters} and \ref{tab:bestfit}). The peak bolometric flux $P_{\gamma}$, in units of $\si{\erg\per\second\per\centi\meter\squared}$, is
\begin{equation}
\label{eq:p_gamma}
P_{\gamma} = L/(4\upi D_L^2).
\end{equation}

Because gamma-ray detectors have limited energy ranges, the observed peak flux is a fraction of the total peak flux. To correct for this, we define a $k$ correction factor from the lab frame to the bolometric rest frame as
\begin{equation}
k=\frac{\int_{1/(1+z)}^{10^4/(1+z)}EN(E)\mathrm{d}E}{\int_{e_{min}}^{e_{max}}EN(E)\mathrm{d}E}
\end{equation}
Here, $e_{max}$ and $e_{min}$ are the maximum and minimum observational energy of the detector. $N(E)$ denotes the photon spectrum of GRBs. We use a typical Band function spectrum \citep{band1993BATSEObservationsGammaray} with $\alpha =-0.5$, $\beta = -2.3$ \citep{preece2000BATSEGammaRayBurst} to simulate the bursts. The peak energy $E_p$ is estimated with the $E_p - E_{iso}$ correlation \citep{amati2002IntrinsicSpectraEnergetics,amati2009ExtremelyEnergeticFermi,kumar2015PhysicsGammarayBursts} but specifically for SGRBs \citep{zhang2012RevisitingLongSoftShort} as:
\begin{equation}
E_p = \SI{2455}{\kilo\eV}~\left(\frac{E_{iso}}{\SI{e52}{\erg}}\right) \left(\frac{0.59}{1 + z}\right).
\end{equation}
The isotropic energy $E_{iso}$ is calculated with $E_{iso} = LT_0$, where the intrinsic duration $T_0$ of SGRBs is drawn from a lognormal distribution with $\mu=-0.3$ and $\sigma=0.57$ \citep{li2016ComparativeStudyLong}.

To convert the observed peak flux $P_{\gamma}/k$ to peak photon flux $P_p$ reported by gamma-ray detectors,
we further introduce a $p$ correction factor, denoting the ratio of energy flux and photon flux as
\begin{equation}
p=\frac{\int_{e_{min}}^{e_{max}}N(E)\mathrm{d}E}{\int_{e_{min}}^{e_{max}}EN(E)\mathrm{d}E},
\end{equation}
then
\begin{equation}
\label{eq:kp_correction}
P_p=\frac{P_{\gamma}}{k}p.
\end{equation}
The simulated peak photon flux $P_p$ is one of the quantities we can directly compare with the observational data.

Because the sensitivity of gamma-ray detectors are limited, only a fraction of all the simulated SGRBs that are bright enough can be ``observed''. Therefore, we must also select SGRBs in our simulation according to their detectability. For this study, we apply a photon flux threshold in line with the detection threshold of the gamma-ray detectors. Considering the non-uniform detection thresholds related to the relative position of the SGRB with respect to the detector, there is a ``grey zone'' inside which the detection sensitivity does not reach a full capacity. We introduce an empirical soft detection probability function to mimic this effect. The detection probability function is a logistic function $1/{1+\exp[-s(P_p-b)]}$ with bias factor $b=0.75$ and scale factor $s=16$. We also set a hard detection limit of $\SI{0.5}{\photon\per\centi\meter\squared\per\second}$ according to the Fermi-GBM sensitivity \citep{vonkienlin2020FourthFermigbmGammaray}

We confront our simulations with two samples. The first sample is the large SGRB sample detected by Fermi-GBM, whose redshifts are largely not measured. For this sample, we mainly compare the simulation results against the observed peak flux distribution, i.e. $\log N-\log P$, which is a convolution of the luminosity function (related to jet structure) and the redshift distribution (related to delay time distribution). 

The second sample is a smaller sample of SGRBs whose redshifts have been measured. These SGRBs are mostly Swift GRBs, but also include some Fermi GRBs.  For these SGRBs, one can use equations \ref{eq:p_gamma}--\ref{eq:kp_correction} and the $k$ and $p$ correction factors described therein in reverse to estimate isotropic luminosities. One can then compare the simulated bursts and the observed bursts in the $z-L$ two-dimensional plane, which carries additional information not available from the $z$-unknown sample.

\begin{figure}
	\includegraphics[width=0.49\textwidth]{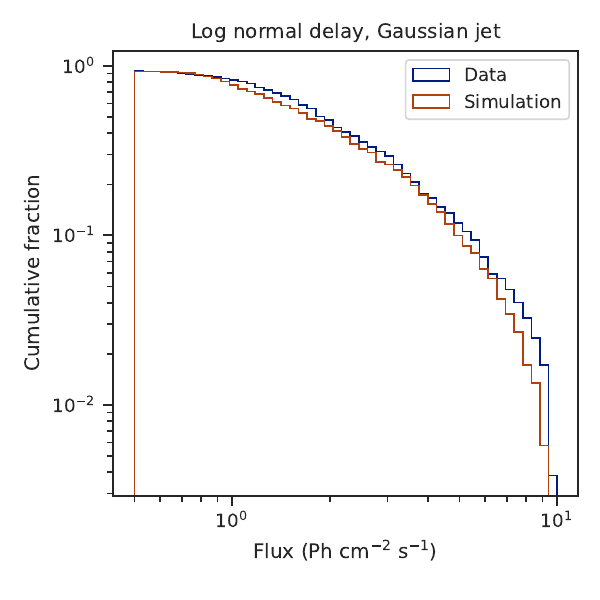}
	\includegraphics[width=0.49\textwidth]{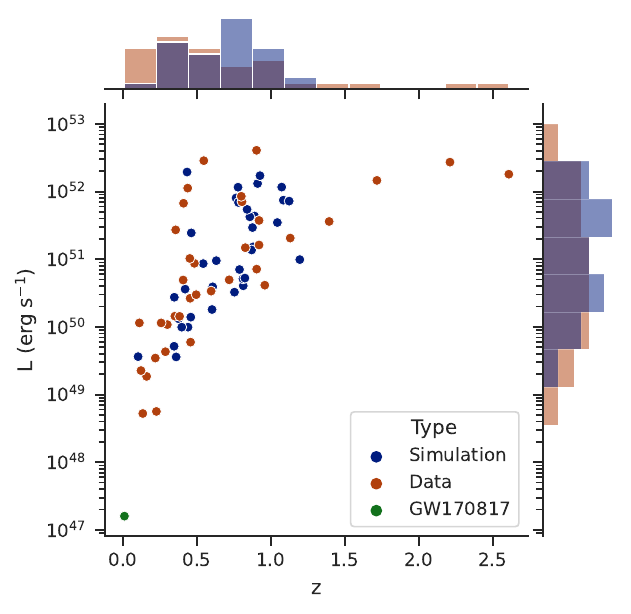}
	\caption{An example of the $\log N-\log P$ plot and the 2D $z-L$ plot for the lognormal delay. Gaussian jet model.}
	\label{fig:example_2d}
\end{figure}

Fig. \ref{fig:example_2d} shows an example of our testing results. Given the same set of model (a delay time model plus a jet structure model), one can utilize two plots, a $\log N-\log P$ plot and a $z-L$ plot to compare the model against the data. For the 2D plot, one can also compare the histograms for each dimension, as shown in the upper and right sides of the plot.

To test how well our simulations reproduce the observed parameter distribution, we utilize the two-sample Kolmogorov–Smirnov test \citep{massey1951KolmogorovSmirnovTestGoodness}. The two-sample KS test compares the cumulative distribution histograms of the observed and simulation data. The maximum difference of the two histograms are taken as the test statistic:
\begin{equation}
    D = \max_i \left| p(i)-q(i) \right|,
\end{equation}
where $p(i)$ and $q(i)$ are normalized cumulative distributions of the observed and simulated parameters. A significance level can be also calculated following \citet{hodges1958SignificanceProbabilitySmirnov}. In this study, we use the two-sample KS test implemented in \texttt{Scipy} \citep{virtanen2020SciPyFundamentalAlgorithms}.

While the two-sample KS test can be directly applied to the $\log N-\log P$ plot, an extension of this test to 2 dimensional data is needed to test the 2D $z-L$ plot. \citep{peacock1983TwodimensionalGoodnessoffitTesting,fasano1987MultidimensionalVersionKolmogorov,press1988KolmogorovSmirnovTestTwoDimensional,lopes2007TwodimensionalKolmogorovSmirnovTest,xiao2017FastAlgorithmTwodimensional}. In this study, we use a modified version of the 2D KS test used by \citet{zhang2021EnergyRedshiftDistributions}.
\begin{equation}
\begin{aligned}
    D = \max_i [\max(&\left| p_1(i)-q_1(i) \right|, \left|p_2(i)-q_2(i)\right|, \\
    &\left|p_3(i)-q_3(i)\right|, \left|p_4(i)-q_4(i)\right|)],
\end{aligned}
\end{equation}
where $p_1(i)$ to $p_4(i)$ and $q_1(i)$ to $q_4(i)$ are the normalized cumulative distribution functions of the observed and simulated data points at the $i$th point defined from the four quadrants, respectively.

\section{Results}
\label{sec:results}

\begin{table*}
	
	\begin{tabular}{ccl}
		\hline
		Model & Parameter & Values\\
		\hline
		\multirow{2}{*}{Gaussian delay}   & $t_G\; (\si{\giga\year})$               & 1.0, 1.5, 2.0, 2.5, 3.0, 3.5, 4.0\\
		                                  & $\sigma_G\; (\si{\giga\year})$          & 0.1, 0.2, 0.3, 0.4, 0.5\\
		\hline
		\multirow{1}{*}{Power-law delay}  & $\alpha$ & 0.7, 0.75, 0.8, 0.85, 0.9\\
		\hline
		\multirow{2}{*}{Lognormal delay} & $t_{LN}\; (\si{\giga\year})$            & 2.5, 3.0, 3.5, 4.0, 4.5\\
		                                  & $\sigma_{LN}\; (\ln{\si{\giga\year}})$  & 0.05, 0.1, 0.15, 0.2\\
		\hline
		\multirow{3}{*}{Gaussian jet}     & $L_0\; (\si{\erg\per\second})$    & logspace(50, 53, num=20)\\
		                                  & \multirow{2}{*}{$\theta_0$} & 0.02, 0.03, 0.04, 0.05, 0.06, 0.07, 0.08, 0.09, 0.10, 0.11, \\
		                                  &                           & 0.12, 0.13, 0.14, 0.15, 0.16, 0.17, 0.18, 0.19, 0.20\\
		\hline
		\multirow{3}{*}{Power-law jet}    & $L_0\; (\si{\erg\per\second})$    & logspace(50, 53, num=10)\\
		                                  & $\theta_0$                              & 0.005, 0.01, 0.015, 0.02\\
		                                  & k                                       & 1.0, 1.3, 1.6, 1.9, 2.2, 2.5, 2.8, 3.1, 3.4, 3.7, 4.0\\
		\hline
		\multirow{4}{*}{Jet+cocoon jet}  & $L_0\; (\si{\erg\per\second})$     & logspace(50, 53, num=10)\\
	                  	                 & $\theta_0$                               & 0.02, 0.04, 0.06, 0.08, 0.10, 0.12\\
		                                 & $L_1\; (\si{\erg\per\second})$     & logspace(48, 51, num=10)\\
		                                 & $\theta_1$                               & 0.15, 0.20, 0.25, 0.30, 0.35, 0.40, 0.45, 0.50\\
		\hline
	\end{tabular}
	\caption{List of model parameters used in simulation. The luminosities shown in this table are the equivalent isotropic luminosities. For Jet+cocoon jet model, $L_1$ and $\theta_0$ are set to be smaller than $L_0$ and $\theta_1$ respectively.}
	\label{tab:parameters}
\end{table*}

We gather SGRB data from the Fermi GRB catalog (\citet{vonkienlin2020FourthFermigbmGammaray}) and filter out the short GRBs with $T_{90}<\SI{2}{\second}$. There are 522 SGRBs in our sample.

We then match the Fermi SGRBs with redshift known SGRBs in the Greiner’s GRB catalog (\citet{greiner2022GRBsLocalizedWithBSAX}) to obtain a list of redshift known Fermi SGRB. For the SGRBs with redshift reported in Greiner’s GRB catalog but are not reported in the Fermi catalog, we complement with the Swift/GRB catalog (\citet{lien2016ThirdSwiftBurst}). We also add GRB170817A to this redshift-known sample. This leaves us a sample of 37 $z$-known SGRBs. The isotropic luminosity of the $z$-known SGRBs are estimated with the method described in Section \ref{sec:test_models_with_data} in reverse. We use $\SIrange{10}{1000}{\kilo\eV}$ as the detector energy range for Fermi \citep{vonkienlin2020FourthFermigbmGammaray}, and $\SIrange{15}{150}{\kilo\eV}$ for Swift \citep{barthelmy2005BurstAlertTelescope}.

For each possible model and parameter combination listed in Table \ref{tab:parameters}, we first test the jet model to ensure that it is enclosed by the constraint of jet parameters from GW 170817 as shown in Fig. \ref{fig:jetmodel}.

To generate more usable data in the same simulation size, we set a minimum isotropic luminosity of $\SI{1e30}{\erg\per\second}$. A maximum viewing angle $\theta$ limit is calculated with the jet model parameters. We only simulate viewing angles smaller than this limit.

Subsequently, we simulate SGRB events in batches of 10000 and apply a sensitivity model with an empirical grey zone. After the detection filtering, we count the number of simulated GRBs that can be detected. We keep simulating batches until we have more than the observed number of GRBs. Finally, we randomly select the same number (522) of GRBs from the simulated sample. We also further randomly select the same number (37) of $z$-known sample. This sub-sampling process makes sure that we have the same number of simulated data points across the parameter space, so it is more straightforward in comparing the goodness-of-fit for each model and parameter combination.

Finally, we compare the simulated $\log N-\log P$ and 2D $z-L$ distributions with the observed distributions as described in Section \ref{sec:method} and obtain test statistics $D_{NP}$ and $D_{zL}$ and the corresponding p-values $p_{NP}$ and $p_{zL}$. We multiply the two significance levels $p_{NP}$ and $p_{zL}$ to obtain a global p-value $p_{global}$. The parameter set with highest global p-value is deemed as the best-fit parameter for each delay and jet model combination.

\begin{figure*}
	\includegraphics[width=0.99\textwidth]{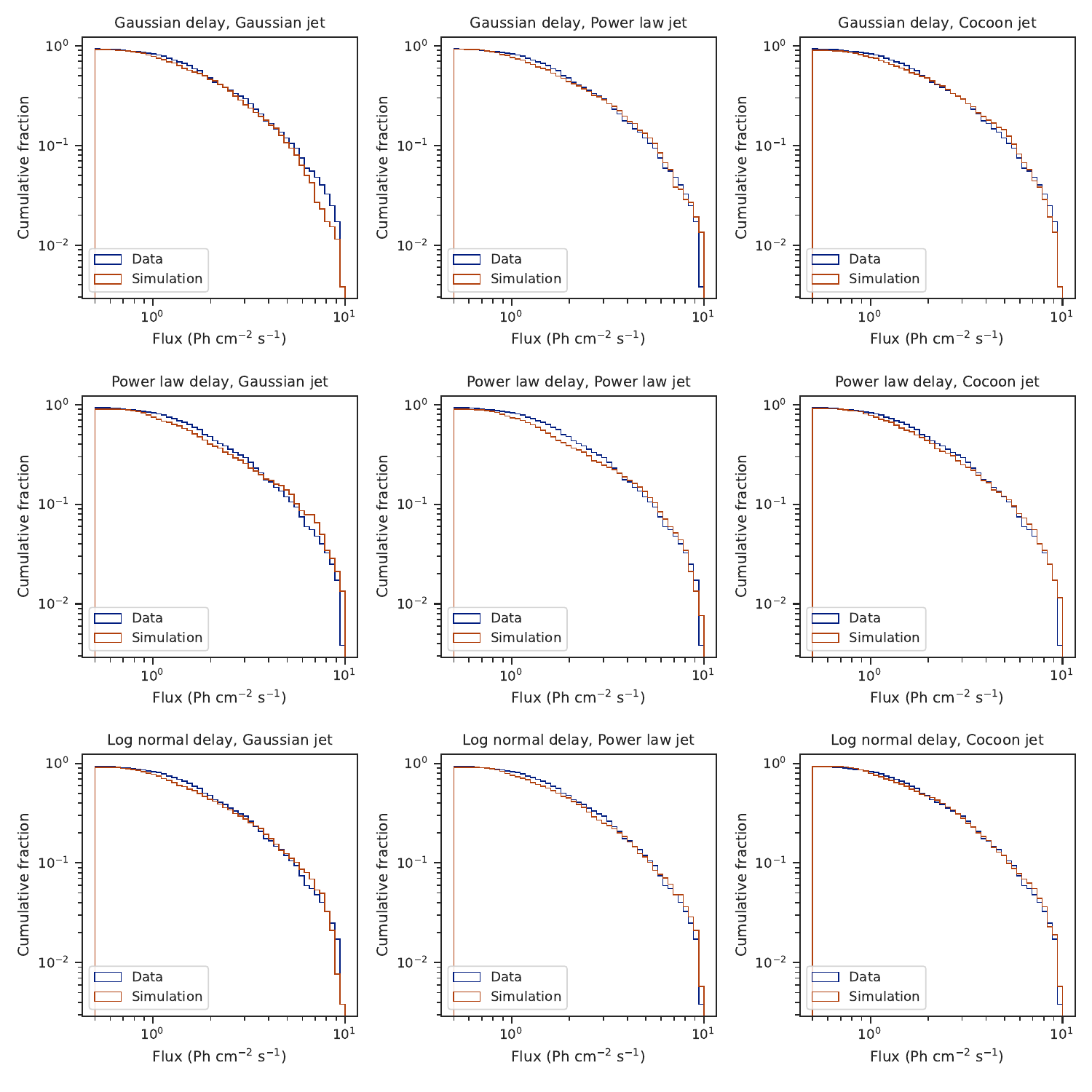}
	\caption{Best fit $\log N-\log P$ distribution for every jet+delay combination model.}
	\label{fig:bestfit_NP}
\end{figure*}

\begin{figure*}
	\includegraphics[width=0.99\textwidth]{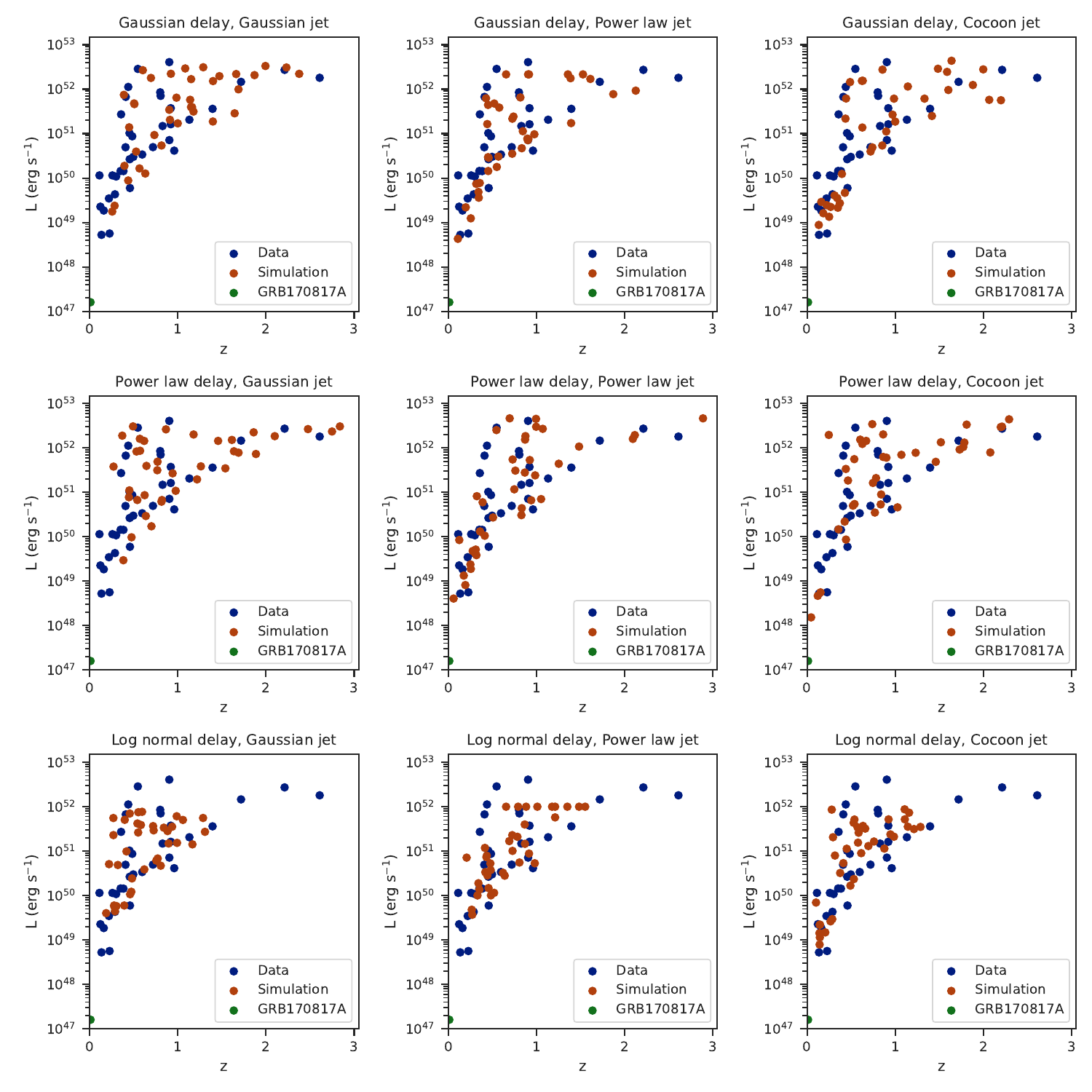}
	\caption{Best fit 2D $z-L$ distribution for every jet+delay combination model.}
	\label{fig:bestfit_zl}
\end{figure*}

The best-fit results are shown in Table \ref{tab:bestfit}, Figure \ref{fig:bestfit_NP}, and Figure \ref{fig:bestfit_zl}. Out of the 9 models we have tested, 5 models are consistent with the data. In particular, the lognormal delay with cocoon jet model performs the best and the Gaussian delay with power-law jet model comes second. On the other hand, 4 models, including all power-law delay time models and the Gaussian delay + Gaussian jet models have $p_{global} < 0.05$, and hence, are disfavored. We also test a positive power-law index in the power-law delay model (negative $\alpha$ in the nomenclature of this study). We find that while such a model can reproduce the $\log N-\log P$ distribution, it fails to reproduce the $z-L$ distribution, and hence, is also disfavored.

\begin{table*}
    \addtolength{\tabcolsep}{-2pt}
	\caption{List of best-fit test statistics and significance for each model combination.}
	\label{tab:bestfit}
	\begin{tabular}{cccccccccc}
		\hline
		Delay model & Jet model & Delay parameter & Jet parameter & $D_{NP}$ & $p_{NP}$ & $D_{zL}$ & $p_{zL}$ & $p_{global}$ & Judgment\\
		\hline
        Gaussian & Gaussian & $t_G=3.0, \sigma_G=0.1$ & $L_0=\num{3.36e52}, \theta_0=0.12$ & 0.044 & 0.692 & 0.405 & 0.009 & 0.006 & \xmark\\
        Gaussian & Power-law & $t_G=4.0, \sigma_G=0.3$ & $L_0=\num{2.15e52}, \theta_0=0.02, k=3.7$ & 0.059 & 0.316 & 0.23 & 0.354 & 0.112 & \cmark\\
        \multirow{2}{*}{Gaussian} & \multirow{2}{*}{Cocoon} & \multirow{2}{*}{$t_G=4.0, \sigma_G=0.5$} & $L_0=\num{4.64e52}, \theta_0=0.02,$ & \multirow{2}{*}{0.067} & \multirow{2}{*}{0.191} & \multirow{2}{*}{0.243} & \multirow{2}{*}{0.288} & \multirow{2}{*}{0.055} & \multirow{2}{*}{\cmark}\\
                                                                     &&& $L_1=\num{4.64e49}, \theta_1=0.15$ &&&&&&\\
        Power-law & Gaussian & $\alpha=0.8$ & $L_0=\num{3.36e52}, \theta_0=0.09$ & 0.067 & 0.191 & 0.392 & 0.013 & 0.003 & \xmark\\
        Power-law & Power-law & $\alpha=0.85$ & $L_0=\num{4.64e52}, \theta_0=0.02, k=4.0$ & 0.084 & 0.049 & 0.189 & 0.601 & 0.029 & \xmark\\
        \multirow{2}{*}{Power-law} & \multirow{2}{*}{Cocoon} & \multirow{2}{*}{$\alpha=0.8$} & $L_0=\num{4.64e52}, \theta_0=0.02,$ & \multirow{2}{*}{0.056} & \multirow{2}{*}{0.396} & \multirow{2}{*}{0.338} & \multirow{2}{*}{0.047} & \multirow{2}{*}{0.019} & \multirow{2}{*}{\xmark}\\
                                                           &&& $L_1=\num{1e49}, \theta_1=0.15$ &&&&&&\\
        Lognormal & Gaussian & $t_{LN}=3.5, \sigma_{LN}=0.1$ & $L_0=\num{7.85e51}, \theta_0=0.10$ & 0.059 & 0.316 & 0.27 & 0.187 & 0.059 & \cmark\\
        Lognormal & Power-law & $t_{LN}=3.0, \sigma_{LN}=0.2$ & $L_0=\num{1e52}, \theta_0=0.02, k=4.0$ & 0.054 & 0.441 & 0.27 & 0.178 & 0.078 & \cmark\\
        \multirow{2}{*}{Lognormal} & \multirow{2}{*}{Cocoon} & \multirow{2}{*}{$t_{LN}=3.5, \sigma_{LN}=0.2$} & $L_0=\num{1e52}, \theta_0=0.04,$  &  \multirow{2}{*}{0.044} & \multirow{2}{*}{0.692} & \multirow{2}{*}{0.216} & \multirow{2}{*}{0.428} & \multirow{2}{*}{0.296} & \multirow{2}{*}{\cmark}\\
                                                                           &&& $L_1=\num{4.64e49}, \theta_1=0.15$ &&&&&&\\
		\hline
	\end{tabular}
	\addtolength{\tabcolsep}{2pt}
\end{table*}

\section{Conclusions and discussion}
\label{sec:conclusions}

In this paper, we have systematically tested the quasi-universal structured jet idea for SGRBs by confronting 9 different delay time and structured jet combinations with the observations including the observed flux, luminosity and redshift distributions. We come up with three conclusions:

\begin{itemize}
    \item The SGRB data is consistent with a quasi-universal jet structure model that interpret all SGRBs and GRB 170817A. The latter is simply a large-viewing-angle event with respect to a standard jet.
    \item The jet structure of this universal model can be loosely constrained. Our results favor the cocoon jet model, whilst the Gaussian and power-law jet models are plausible in some cases. However, the Gaussian jet model with a Gaussian delay time distribution is disfavored by the data.
    \item We can also constrain the merger delay time distribution model alongside with the jet model under the quasi-universal jet hypothesis. The lognormal delay time distribution model achieves the best result, and the Gaussian delay model follows behind. The power-law delay models (including both negative and positive slopes) are completely disfavored in our model combination and parameter space.
\end{itemize}

There is an important caveat that our study is based on the assumption that all SGRBs share the same set of delay time distribution, jet profile, and model parameters. While the models that work under these assumptions will still work if there are variations of jet model and parameters across different SGRBs, the disfavored ones might perform better under real-world circumstances. 

We also did not make use of other observational criteria such as the host galaxy properties of SGRBs. For example, \cite{zheng2007DeducingLifetimeShort} and recently \cite{zevin2022ObservationalInferenceDelay} argued that a power-law delay time model (which is disfavored by our analysis) is consistent with the host galaxy data. Since these authors did not test the lognormal or Gaussian delay models, it remains unclear whether there is a direct conflict between the two claims. If the lognormal and Gaussian delay models cannot pass the host galaxy constraints, then there is a direct conflict and one may need to drop the hypothesis that all SGRBs share a similar quasi-universal jet structure. More joint GW-SGRB events to be detected in the future can pose important constraints on the hypothesis and the tested jet and delay models.

\section*{Acknowledgements}
This work is partially supported by the Top Tier Doctoral Graduate Research Assistantship (TTDGRA) and Nevada Center for Astrophysics at the University of Nevada, Las Vegas. YL is partially supported by the Natural Science Foundation of China (Grant Nos. 12041306, 12103089), the Natural Science Foundation of Jiangsu Province (Grant No. BK20211000) and China Manned Spaced Project (CMS-CSST-2021-B11). HG is supported by the National Natural Science Foundation of China (Projects:12021003).

\section*{Data availability}
\label{sec:data_availability}
The data used in this paper are public and are available in corresponding references. The code can be shared upon request to the authors.

\bibliographystyle{mnras}
\bibliography{grb-jet}
\bsp
\label{lastpage}
\end{document}